\documentclass[twoside,12pt]{article}
\usepackage{epsfig}

\def\Journal#1#2#3#4{{#1} {#2} (#4) #3 }

\def\NPA{{\em Nucl. Phys.} A}

\def\PLB{{\em Phys. Lett.} B}

\def\PRC{{\em Phys. Rev.} C}

\newcommand{\be}{\begin{equation}}
\newcommand{\ee}{\end{equation}}
\newcommand{\bea}{\begin{eqnarray}}
\newcommand{\eea}{\end{eqnarray}}

\begin{document}

\title{ \vspace{1cm} Effect of deformation on two-neutrino double beta
decay matrix elements}

\author{R.\ \'Alvarez-Rodr\'{\i}guez,$^{1}$ P.\ Sarriguren,$^1$ E.\ Moya
de Guerra,$^1$ \\ 
L.\ Pacearescu,$^{2}$  A.\ Faessler,$^{2}$ F.\ \v Simkovic$^{2,3}$\\ 
\\
$^1$Instituto de Estructura de la Materia, Consejo Superior de \\
Investigaciones Cient\'{\i}ficas, Serrano 123, E-28006 Madrid, Spain\\
$^2$Institut f\"ur Theoretische Physik, Universit\"at T\"ubingen, \\
D-72076 T\"ubingen, Germany\\
$^3$Department of Nuclear Physics, Comenius University, \\SK-842 15 
Bratislava, Slovakia}
\maketitle

\begin{abstract} 
We study the effect of deformation on the two-neutrino double beta decay
for ground state to ground state transitions in all the nuclei whose 
half-lives have been measured. Our theoretical framework is a deformed
QRPA based in Woods-Saxon or Hartree-Fock mean fields. We are able to
reproduce at the same time the main characteristics of the two single 
beta branches, as well as the double beta matrix elements. We find a
suppression of the double beta matrix element with respect to the 
spherical case when the parent and daughter nuclei have different
deformations. 
\end{abstract}

It has been observed that the matrix elements $M_{2\nu}$ in all the
measured $2\nu\:\beta\beta$ cases for the $0^+ \to 0^+$ transition are
quenched with respect to those predicted from pure quasiparticle
transitions. The physical mechanisms responsible for this reduction
have been a subject of great interest. In this work, we study the effect
of the parent and daughter deformations on the $M_{2\nu}$ matrix elements.

We use a deformaed quasiparticle random phase approximation (pnQRPA) to
describe simultaneously the energy distributions of the single $\beta$
Gamow-Teller (GT) strength and the matrix element $M_{2\nu}$ of all the
double beta emitters whose half-lives have been measured: $^{48}$Ca, 
$^{76}$Ge, $^{82}$Se, $^{96}$Zr, $^{100}$Mo, $^{116}$Cd, $^{128}$Te, 
$^{130}$Te, $^{136}$Xe, and $^{150}$Nd. The formalism includes a deformed
quasiparticle basis and residual spin-isospin separable interactions in
both particle-hole (ph) and particle-particle (pp) channels. We consider
two different deformed mean fields: a phenomenological Woods-Saxon (WS)
potential and a selfconsistent Hartree-Fock (HF) with Skyrme interactions
\cite{alv1}. While the deformation in the HF calculation is obtained
selfconsistently, in the WS case it is an input parameter taken to reproduce
the experimental quadrupole moments.

We describe the $2\nu\:\beta\beta$ process as two successive GT
transitions via intermediate $1^+$ states,
\begin{equation}
 M^{2\nu}_{\rm GT}=\sum_{K=0,\pm 1}\sum_{m_i,m_f}
\frac{\left\langle 0^+_f \left| \right| \sigma_K t^- \left| \right|
\omega_K^{m_f}\right\rangle
\left\langle \omega_K^{m_f} \left| \right. \omega_K^{m_i} \right\rangle
\left\langle \omega_K^{m_i} \left| \right| \sigma_K t^- \left| \right|
0^+_i \right\rangle }
{\left( \omega_K^{m_f}+\omega_K^{m_i}\right) /2 }\;.
\end{equation}
The overlap is necessary to take into account the non-orthogonality of the
intermediate $1^+$ states reached from both parent and daughter nuclei
\cite{simko1}.

We study first the single GT strength distributions and compare them to the
experimental data available. Then, we study the sensitivity  of the matrix
elements $M_{2\nu}^{GT}$ to the nuclear deformation and to the pp strength,
which is known to be a suppression mechanism. In the case of HF we reproduce
simultaneously this information using coupling strengths 
$\chi_{GT}^{ph} = 0.1$ MeV and $\kappa_{GT}^{pp}=6/A$ MeV, while in the case 
of WS we find that the parametrization of Homma \cite{homma}, 
$\chi_{GT}^{ph}=5.2/A^{0.7}$ MeV and $\kappa_{GT}^{pp}= 0.58/A^{0.7}$ MeV,
reproduces better the experiment. As an example, 
we show in Fig. 1 the results for
$^{100}$Mo. We notice that deformation introduces a reduction factor with
respect to the spherical result. This reduction is larger as the difference
between the deformation of parent and daughter increases.

This result is an important starting point to test the validity of this 
kind of nuclear structure calculations and to tackle the $0\nu\:\beta\beta$
decay. 

\begin{figure}[tb]
\begin{center}
\begin{minipage}[t]{15 cm}
\centering
\epsfig{file=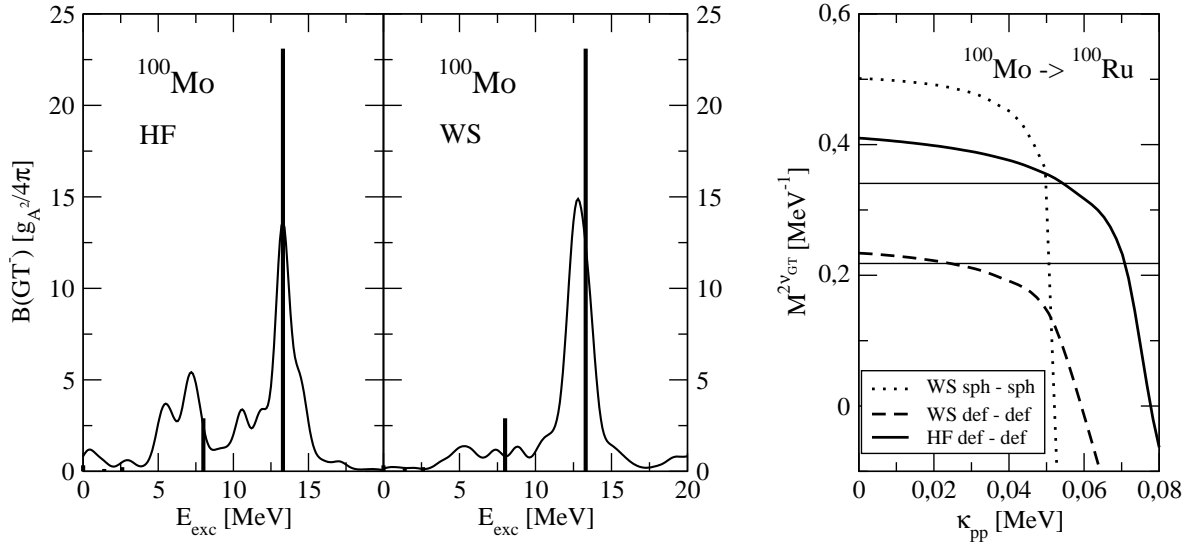,scale=0.58,angle=-90}
\end{minipage}
\begin{minipage}[t]{16.5 cm}
\caption{Left panel: GT$^-$ strength distributions in $^{100}$Mo using HF
and WS mean fields. Data (vertical lines) are from Ref. \cite{akim}. 
Right panel: $2\nu\:\beta\beta$-decay matrix element as a function of the
pp interaction strength. Horizontal lines are the experimental 
$M_{GT}^{2\nu}$ values extracted from Ref. \cite{bara} using $g_A=1.0$ and 
$g_A=1.25$. }
\end{minipage}
\end{center}
\label{figure1}
\end{figure}
\vskip 0.5cm
This work was supported in part by MEC (Spain) under contract number 
BFM2002-03562. One of us (R.A.R.) thanks Ministerio de Educaci\'on y
Ciencia (Spain) for financial support.

\end{document}